# High-order fundamental and general solutions of convection-diffusion equation and their applications with boundary particle method

W. Chen[*]

Simula Research Laboratory, P. O. Box. 134, 1325 Lysaker, Norway

E-mail: wenc@simula.no

**Abstract**

In this study, we presented the high-order fundamental solutions and general solutions of convection-diffusion equation. To demonstrate their efficacy, we applied the high-order general solutions to the boundary particle method (BPM) for the solution of some inhomogeneous convection-diffusion problems, where the BPM is a new truly boundary-only meshfree collocation method based on multiple reciprocity principle. For the sake of completeness, the BPM is also briefly described here.

## 1. Introduction

The high-order fundamental solutions play a pivotal role in the multiple reciprocity BEM (MR-BEM) [1]. On the other hand, the boundary particle method (BPM), recently introduced by the present author [2], also requires the use of the high-order general solutions. In addition, an underlying relationship between the radial basis function (RBF) and Green second identity highlights the importance of these operator-dependent solutions [3]. A kernel RBF-creating strategy [3,4], which employs these high-order fundamental or general solutions as the RBF, is presented for the dual reciprocity BEM

---

[*] This research is supported by Norwegian Research Council.

[5], boundary knot method [6], and Kansa's method [7].

To the author's best knowledge, the explicit high-order fundamental solutions are now available only for the Laplace [1], Helmholtz and modified Helmholtz operators [8]. Based on the underlying relationship between the convection-diffusion equation and the modified Helmholtz equation [9], we derive the high-order fundamental and general solutions of the convection-diffusion equation [10]. In this study, we will give a systematic description of this recent advance. In addition, to demonstrate the efficacy of these high-order solutions, we apply them to the boundary particle method for the solution of some inhomogeneous convection-diffusion problems.

The BPM is developed based on the multiple reciprocity principle [2,10], which uses either high-order nonsingular general solutions or singular fundamental solutions to evaluate high-order homogeneous solutions, and then their sum approximates the particular solutions. Like the MR-BEM, the BPM does not require any inner nodes for inhomogeneous problems and therefore is a truly boundary-only technique. Unlike the MR-BEM, the BPM does not need to generate more than one interpolation matrix since we developed the efficient recursive multiple reciprocity formulas, which tremendously reduces computing effort and memory requirements. In addition, the BPM is essentially meshfree, spectral convergence, easy-to-use, and symmetric technique.

The rest of this paper is organized into 4 sections. We discuss the difference between the singular fundamental solution and nonsingular general solution and then introduce the high-order fundamental and general solutions of the convection-diffusion equation in section 2. The boundary particle method is briefly described in section 3, and followed by the implementation issues: overflow and ill conditioning in section 4. Finally, Section 5 presented numerical testing the convection-diffusion high-order general solutions with the BPM to some 2D inhomogeneous convection-diffusion problems under complicated geometries.

## 2. High-order fundamental and general solutions of Convection-diffusion operator

The general solution $u^{\#}$ satisfies

$$\Re\{u^{\#}\} = 0, \tag{1}$$

where $\Re$ is a differential operator. The general solution at origin is equal to a limit value rather than zero and infinity. In contrast, the fundamental solution $u^{*}$ has to satisfy

$$\Re\{u^{*}\} + \Delta_i = 0, \tag{2}$$

where $\Delta_i$ represents the Dirac delta function which goes to infinity at the origin point $x_i$ and is equal to zero elsewhere. According to the above definitions, the essential distinctness between the general solution and fundamental solution of a differential operator is that the former is nonsingular while the latter is singular.

### 2.1. High-order general solutions

Consider a homogeneous isotropic region $\Omega$. The steady convection-diffusion equation is given by

$$D\nabla^2 u(x) - \vec{v} \bullet \nabla u(x) - \kappa u(x) = 0, \quad \text{in } \Omega \tag{3}$$

where $\vec{v}$ denotes a velocity vector, $D$ is the diffusivity coefficient, $\kappa$ represents the reaction coefficient. The variable $u$ can be interpreted as temperature for heat transfer problems, concentration for dispersion problems [5].

Assuming that convective velocity $\vec{v}$ and diffusivity $D$ are constant, we have an exponential variable transformation [9]

$$u(x) = \exp\left[\frac{\vec{v} \bullet \vec{r}}{2D}\right] w(x), \tag{4}$$

where $w(x)$ is an intermediate field variable, and $\vec{r}$ is the distance vector between the source and field points. Convection-diffusion equation (3) can be accordingly rewritten as a modified Helmholtz equation

$$\nabla^2 w - \tau^2 w = 0 \quad \text{in } \Omega, \tag{5}$$

where

$$\tau = \left[\left(|\vec{v}|/2D\right)^2 + \kappa/D\right]^{\frac{1}{2}}. \tag{6}$$

According to Itagaki's three-dimensional high-order fundamental solutions of the modified Helmholtz equation [8], it is very easy to get the corresponding higher-order general solutions, i.e.

$$w_m^{\#}(r) = A_m (\tau r)^{-n/2+1+m} I_{n/2-1+m}(\tau r), \quad n \geq 2, \tag{7}$$

where $A_m = A_{m-1}/(2m\tau^2)$, $A_0 = 1$; $n$ is the dimension of the problem; $m$ denotes the order of general solution; $r$ is the Euclidean distance norm; $I$ represents the modified Bessel function of the first kind, which can be transformed to the modified spherical Bessel function when $n=3$. Using computer software "Maple", we verified that those higher-order fundamental and general solutions are also established for more than 3 dimensions.

Through an inverse transformation process, we derive the high-order general solutions of the convection-diffusion equation

$$u_m^{\#}(r) = A_m e^{\frac{\vec{v} \cdot \vec{r}}{2D}} (\tau r)^{-n/2+1+m} I_{n/2-1+m}(\tau r), \quad n \geq 2, \tag{8}$$

where $\tau$ is defined by formula (6).

The above high-order general solutions of the convection-diffusion equation are also validated by mathematical deduction approach with the help of computer algebraic package "Maple". Namely, first we prove the zero-order general solution, and then, it is found that $\Re\{u_m^\#\}$ consists only of ($m$-1)th or lower order general solutions of operator $\Re$. Therefore, the m-order general solution is proved.

It is noted that the general solutions (7) and (8) are nonsingular at the origin and actually infinitely continuous but grow exponentially as $r \to \infty$. Thus, it is often regarded as "improper" solutions [11]. In contrast, the fundamental solution decays exponentially and is considered as "proper" solution. However, if our problem is to handle the data or PDEs on the bounded domains, i.e. interior problems, the growth at infinity is no longer an issue, since the expansion terms always remain finite [11]. We have some persuasive numerical demonstrations of the utility of these general solutions with the boundary knot method [7,10] in solving interior problems. In later section, we will show these general solutions in conjunction with the BPM could be very successfully applied to evaluation of some typical convection-diffusion problems on bounded domains.

In general, the fundamental solutions can be applied to both external and interior problems, while the general solutions are but the solution of interior problems. But the general solutions may nevertheless be applicable even to exterior problems if we use the truncated domain approach, e.g., for exterior wave problems with a proper nonreflecting boundary condition at truncated boundary.

**2.2. High-order fundamental solutions**

The fundamental solution of steady convection-diffusion equation is determined by

$$D\nabla^2 u(x) - \bar{v} \bullet \nabla u(x) - \kappa u(x) = -\Delta_i, \text{ in } \Omega \qquad (9)$$

where $\Delta_i$ represents the Dirac delta function at a source point $i$. Following the preceding procedure for high-order general solutions, we can easily derive the m-order fundamental solution

$$u_m^*(r) = B_m e^{\frac{\vec{v}\cdot\vec{r}}{2D}} (\sigma r)^{-n/2+1+m} K_{n/2-1+m}(\sigma r), n \geq 2, \quad (10)$$

where $K$ represents the modified Bessel function of the second kind. As in the modified Helmholtz case, the high-order fundamental solution of convection-diffusion equation has the singularity order of ($r^{2-n}$) except the 2D case where the only singularity occurs in the zero-order fundamental solution. For instance, the singularity for 3D case is always $r^{-1}$ irrespective of the order of fundamental solution. In other words, the singularity order has nothing to do with the order of a fundamental solution.

### 2.3. Evaluating fundamental and general solutions

It is worth pointing out that we can greatly simplify the above-given standard forms of high-order solutions involving some Bessel functions. Thus, the computing effort for them could be trivial. For example, the zero-, first-, and second-order general solutions of 3D convection-diffusion equation are respectively simplified via computer algebra package "Maple" as

$$u_0^\#(r) = \frac{A_0 \sqrt{2}}{\sqrt{\pi}} e^{\frac{\vec{v}\cdot\vec{r}}{2D}} \frac{\sinh(\sigma r)}{\sigma r}, \quad (11)$$

$$u_1^\#(r) = \frac{A_1 \sqrt{2}}{\sqrt{\pi}} e^{\frac{\vec{v}\cdot\vec{r}}{2D}} \frac{r\cosh(\sigma r) - \sinh(\sigma r)}{\sigma r}, \quad (12)$$

$$u_2^\#(r) = \frac{A_2 \sqrt{2}}{\sqrt{\pi}} e^{\frac{\vec{v}\cdot\vec{r}}{2D}} \frac{r^2 \sinh(\sigma r) + 3\sinh(\sigma r) - 3r\cosh(\sigma r)}{\sigma r}, \quad (13)$$

where sinh and cosh respectively denote the sine-hyperbolic and cosine-hyperbolic functions.

## 3. Boundary particle methods

In section 4 the above-derived high-order general solutions of convection-diffusion equation are applied with the boundary particle method to some 2D inhomogeneous problems. For the completeness and very recent origin of the BPM, we give a brief description of the method in this section. Consider a steady convection-diffusion system

$$D\nabla^2 u(x) - \vec{v} \bullet \nabla u(x) - \kappa u(x) = f(x), \qquad x \in \Omega, \tag{14}$$

$$u(x) = R(x), \qquad x \subset S_u, \tag{15}$$

$$\frac{\partial u(x)}{\partial n} = N(x), \qquad x \subset S_T, \tag{16}$$

where $n$ is the unit outward normal. The system solution can be split as

$$u = u_h^0 + u_p^0, \tag{17}$$

where $u_h^0$ and $u_p^0$ are the zero-order homogeneous and particular solutions, respectively. The multiple reciprocity principle assumes that the sufficiently smooth particular solution can be approximated by higher-order homogeneous solutions, i.e.

$$u = u_h^0 + u_p^0 = u_h^0 + \sum_{m=1}^{\infty} u_h^m, \tag{18}$$

where superscript $m$ is the order index of homogeneous solution. Through an incremental differentiation operation via the convection-diffusion operator, we have successively higher order boundary differential equations:

$$\begin{cases} u_h^0(x_i) = R(x_i) - u_p^0(x_i) \\ \dfrac{\partial u_h^0(x_j)}{\partial n} = N(x_j) - \dfrac{\partial u_p^0(x_j)}{\partial n} \end{cases}, \qquad (19a)$$

$$\begin{cases} \Re^0\{u_h^1(x_i)\} = f(x_i) - \Re^0\{u_p^1(x_i)\} \\ \dfrac{\partial \Re^0\{u_h^1(x_j)\}}{\partial n} = \dfrac{\partial (f(x_j) - \Re^0\{u_p^1(x_j)\})}{\partial n} \end{cases}, \qquad (19b)$$

$$\begin{cases} \Re^{m-1}\{u_h^m(x_i)\} = \Re^{m-2}\{f(x_i)\} - \Re^{m-1}\{u_p^m(x_i)\} \\ \dfrac{\partial \Re^{m-1}\{u_h^m(x_j)\}}{\partial n} = \dfrac{\partial (\Re^{m-2}\{f(x_j)\} - \Re^{m-1}\{u_p^m(x_j)\})}{\partial n} \end{cases}, \quad m=2,3,\ldots, \qquad (19c)$$

where $\Re^m\{\}$ denotes the $m$-th order of convection-diffusion equation, say $\Re^1\{\}=\Re\Re^0\{\}$, and $\Re^0\{\}=\Re\{\}$, $i$ and $j$ are knots indices respectively on Dirichlet and Neumann boundary. The m-order homogeneous solution can be approximated in terms of the m-order general solution $u_m^\#$ basis function

$$u_h^m(x) = \sum_{k=1}^{L} \beta_k^m u_m^\#(r_k), \qquad (20)$$

where $L$ is the number of boundary nodes, and $k$ denotes the index of source points on boundary. $\beta_k$ are the desired coefficients. Note that we can use the fundamental solution $u_m^*$ instead of general solution $u_m^\#$ in Eq. (20). However, the use of fundamental solution requires a fictitious boundary outside physical domain as encountered in the MFS [12].

Collocating Eqs. (19a,b,c) at all boundary knots in terms of representation (20), we have the BPM boundary discretization equations :

$$\left.\begin{array}{l}\sum_{k=1}^{L}\beta_k^0 u_0^\#(r_{ik})=R(x_i)-u_p^0(x_i)\\ \sum_{k=1}^{L}\beta_k^0 \dfrac{\partial u_0^\#(r_{jk})}{\partial n}=N(x_j)-\dfrac{\partial u_p^0(x_j)}{\partial n}\end{array}\right\}=b^0, \qquad (21a)$$

$$\left.\begin{array}{l}\sum_{k=1}^{L}\beta_k^1 \Re^0\{u_1^\#(r_{ik})\}=f(x_i)-\Re^0\{u_p^1(x_i)\}\\ \sum_{k=1}^{L}\beta_k^1 \dfrac{\partial \Re^0\{u_1^\#(r_{jk})\}}{\partial n}=\dfrac{\partial(f(x_j)-\Re^0\{u_p^1(x_j)\})}{\partial n}\end{array}\right\}=b^1, \qquad (21b)$$

$$\left.\begin{array}{l}\sum_{k=1}^{L}\beta_k^m \Re^{m-1}\{u_m^\#(r_{ik})\}=\Re^{m-2}\{f(x_i)\}-\Re^{m-1}\{u_p^m(x_i)\}\\ \sum_{k=1}^{L}\beta_k^m \dfrac{\partial \Re^{m-1}\{u_m^\#(r_{ik})\}}{\partial n}=\dfrac{\partial(\Re^{m-2}\{f(x_j)\}-\Re^{m-1}\{u_p^m(x_j)\})}{\partial n}\end{array}\right\}=b^m, \qquad m=2,3,\dots \qquad (21c)$$

Whenever the mixed boundary conditions are present, the present BPM interpolation matrix is unsymmetric even for self-adjoint operator. By analogy with Fasshauer's Hermite RBF interpolation [13] the present author also proposed a symmetric BPM formulation for self-adjoint operator [2]. Since the convection-diffusion equation is not self-adjoint, this study simply uses the present unsymmetric BPM scheme.

As in the MR-BEM [1], the successive process is terminated at some order *M*, namely,

$$\Re^{M-1}\{u_p^M\}=0. \qquad (22)$$

The practical solution procedure is a reversal recursive process:

$$\beta^M \to \beta^{M-1} \to \cdots \to \beta^0. \qquad (23)$$

It is noted that due to

$$\Re^{m-1}\{u_m^\#(r)\}=u_0^\#(r),\tag{24}$$

the coefficient matrices of all successive equations (21a,b,c) are the same, i.e., let

$$Q\beta^m = b^m, \; m=M,M\text{-}1,\ldots,1,0.\tag{25}$$

The LU decomposition algorithm is suitable for this task. Then we can employ the obtained expansion coefficients $\beta$ to calculate the BPM solution at any knot, i.e.

$$u(x_i)=\sum_{m=0}^{M}\sum_{k=1}^{L}\beta_k^m u_m^\#(\|x_i-x_k\|).\tag{26}$$

It is noted that the BPM solution (26) is also a RBF wavelet expansion series, where the differential equation order $m$ is used as the dilation parameter and $x_k$ is the translation parameter. It is also worth pointing out that the above multiple reciprocity solution procedure is in fact equivalent to that within the MR-BEM. The distinction is that the present one has a mathematical procedure computationally more efficient since it is a recursive iteration solution which does not need to evaluate costly interpolation matrices of high order partial differential equations. We will give a detailed discussion on this issue and present the MR-BEM formulations based on the general solution and the fundamental solution in a subsequent paper.

**4. Implementation issues: overflow and ill-conditioning**

As the parameter $\tau$ in formula (6) grows, the condition number of the BPM discretization matrix $Q$ in (25) increases quickly and the maximum and minimum entries respectively associated with the general solution and the fundamental solution become prone to overflow. Consequently, the solution blows up at medium value of $\tau$. With the normal numerical algebraic procedure, the overflow and ill-conditioning are two major bottlenecks in applying the BPM to convection problems with large $\tau$ value.

To avoid the overflow, this study modifies the BPM interpolation matrix $Q$ in (25) by multiplying it with an exponential parameter to scale down the maximum entry with the general solution, i.e.

$$\{e^{-\tau L_c} u_0^\#(r_{ij})\}\{e^{\tau L_c} \beta_j^m\} = b^m, \quad m=M, M-1, \ldots, 1, 0, \tag{27}$$

where $L_c$ denotes a characteristic length. The entries of interpolation matrix in (27) is actually calculated in the following manner:

$$e^{-\tau L_c} u_0^\#(r_{ij}) = A_0 e^{-\tau L_c + \frac{\vec{v}\cdot\vec{r}_{ij}}{2D} + \tau r_{ij}} (\tau r_{ij})^{-n/2+1} \left[e^{-\tau r_{ij}} I_{n/2-1}(\tau r_{ij})\right], \quad n \geq 2. \tag{28}$$

Here $e^{-\tau r_{ij}} I_{n/2-1}(\tau r_{ij})$ is seen as a new function, which is never overflow and much more smooth than the modified Bessel function $I_{n/2-1}(\tau r_{ij})$. Many software packages include the subroutine or special function for calculating $e^{-\tau r_{ij}} I_{n/2-1}(\tau r_{ij})$. It is worth noting that corresponding to (27), the formula (26) is revised as

$$u(x_i) = \sum_{m=0}^{M} \sum_{k=1}^{L} \left(e^{\tau L_c} \beta_k^m\right) \left[e^{-\tau L_c} u_m^\#(\|x_i - x_k\|)\right], \tag{29}$$

where

$$e^{-\tau L_c} u_m^\#(r_{ij}) = A_m e^{-\tau L_c + \frac{\vec{v}\cdot\vec{r}_{ij}}{2D} + \tau r_{ij}} (\tau r)^{-n/2+1+m} \left[e^{-\tau r_{ij}} I_{n/2-1+m}(\tau r_{ij})\right], \quad n \geq 2. \tag{30}$$

If the BPM uses the fundamental solution, we need to scale up the minimum entry in comparison to (28), i.e.

$$e^{\tau L_c}u_0^*(r) = B_0 e^{\tau L_c + \frac{\vec{v}\cdot\vec{r}}{2D} - \tau r}(\tau r)^{-n/2+1}\left[e^{\tau r}K_{n/2-1}(\tau r)\right], \; n \geq 2. \tag{31}$$

The remainder of the procedure is similar to that of the BPM using the general solution.

On the other hand, we still lack a good devise to deal with the ill-conditioning issue. Here we present a conceptual approach called Hadamard product preconditioning.

**Definition 4.1** Let matrices $P=[p_{ij}]$ and $Q=[q_{ij}] \in C^{N \times M}$, the Hadamard product of matrices is defined as $P \circ Q = [p_{ij} q_{ij}] \in C^{N \times M}$, where $C^{N \times M}$ denotes the set of $N \times M$ real matrices.

Observing the BPM interpolation matrix $Q$ in (25), we note that $Q$ can be decomposed via the Hadamard matrix product, i.e.

$$Q = \{e^{\tau r_{ij}}\}\left\{A_0 e^{\frac{\vec{v}\cdot\vec{r}_{ij}}{2D}}(\tau r_{ij})^{-n/2+1}\left[e^{-\tau r_{ij}}I_{n/2-1}(\tau r_{ij})\right]\right\} = H \circ \hat{Q}. \tag{32}$$

Thus, the formulation (25) is restated as

$$[H \circ \hat{Q}]\beta^m = b^m, \; m=M,M\text{-}1,\ldots,1,0. \tag{33}$$

Note that the matrix $\hat{Q}$ will never be overflow and is well conditioned. At issue is how to manipulate and precondition the augmented $H$, which is a poorly-conditioned exponential function matrix. Albeit extremely simple in its form, the Hadamard product is quite hard to analyze and maneuver [14-16]. For the time being, the ill-conditioning is still an open issue in the BPM solution of the convection-diffusion problem.

## 5. Numerical validations

The tested case is a 2D steady convection-diffusion problem described by

$$D\nabla^2 u - \vec{v} \bullet \nabla u - \kappa u = e^{-\eta(x+y)}(2 - 4\eta x + 2\sigma x) \tag{34}$$

with the boundary conditions of the Dirichlet type (15) and Neumann type (16), where $D=1$, $v_x=v_y=-\sigma$, $\kappa=3\sigma^2/2$, $\eta = \left(\sigma + \sqrt{\sigma^2 + 2\kappa}\right)/2$. The exact solution is

$$u = x^2 e^{-\eta(x+y)}. \tag{35}$$

The Peclet number Pe is defined as

$$Pe = \frac{|\vec{v}|L_c}{D}. \tag{36}$$

The Peclet number indicates the relative importance of convection versus diffusion. Fig. 1. illustrates the 2D irregular geometry. Except specified Neumann boundary conditions at $x=0$ and $y=0$ edges, the otherwise boundary are all of the Dirichlet type. Equally spaced knots were applied on the boundary. The $L_2$ norms of relative errors are calculated based on the numerical solutions at 460 inner and boundary nodes. The absolute error is taken as the relative error if the absolute value of the solution is less than 0.001.

Table 1 displays the experimental results, where $L$ represents the number of the used boundary knots. It is seen that the BPM scheme based on the high-order general solutions produces stable solutions with increasing number of boundary nodes for the present inhomogeneous convection-diffusion problems. In particular, the BPM solutions are found very accurate even with a small number of boundary nodes. It is stressed that we did not use any inner nodes here. For higher Peclect number problems, the BPM solutions become unstable due to very poor condition of the BPM interpolation matrix. As in the BEM, some preconditioning techniques could be used to handle this issue, which is beyond the present study. Now one can conclude that the presented high-order general solutions of convection-diffusion equation do work well in this numerical experiment.

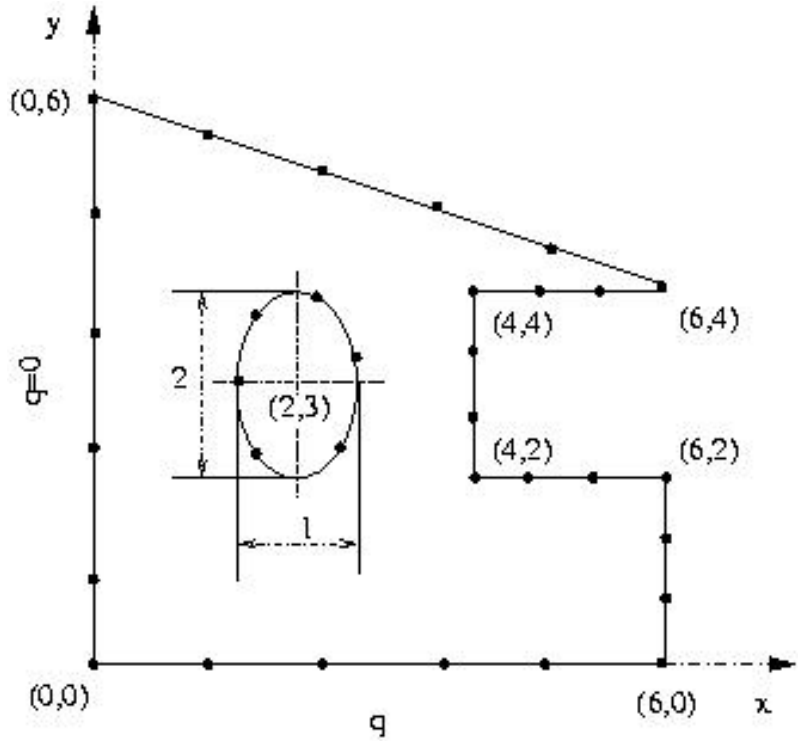

Fig. 1. Configuration of 2D irregular geometry

Table 1. $L_2$ norm of relative errors for 2D inhomogeneous convection-diffusion problem.

| L | 25 | 33 | 41 | 49 |
|---|---|---|---|---|
| Pe=24 | 9.0e-3 | 1.4e-3 | 3.6e-4 | 3.6e-4 |
| L | 33 | 41 | 49 | 57 |
| Pe=72 | 1.2e-2 | 4.5e-2 | 2.2e-3 | 4.5e-4 |